\documentclass[twocolumn,english,aps,prb,floatfix]{revtex4}
\usepackage[T1]{fontenc}
\usepackage[latin2]{inputenc}
\usepackage{amsmath}
\usepackage{graphicx}
\usepackage{amssymb}

\makeatletter
\usepackage{bm}

\usepackage{babel}
\makeatother
\begin{document}

\title{Conductance of interacting Aharonov-Bohm systems}

\author{T. Rejec$^{1}$ and A. Ram\v{s}ak$^{1,2}$}

\affiliation{$^{1}$Jo\v{z}ef Stefan Institute, Jamova 39, SI-1000 Ljubljana, Slovenia\\
 $^{2}$Faculty of Mathematics and Physics, University of Ljubljana,
Jadranska 19, SI-1000 Ljubljana, Slovenia }

\begin{abstract}
A simple formula for the zero-temperature linear response conductance
of an interacting mesoscopic region, threaded by magnetic flux, and
attached to noninteracting single-channel leads is presented. The
formula is valid for a general interacting system exhibiting Fermi
liquid properties. As an example of the efficiency of the formula
the results for the conductance of a simple Aharonov-Bohm ring with
Kondo-Fano resonance physics are presented and compared with numerical
renormalization group results.
\end{abstract}


\date{13 March 2003}

\maketitle
The electron-electron interaction often plays a crucial role in transport
through mesoscopic systems. The Kondo effect in quantum dots is e.g.
a prototype phenomenon, where correlations dominate the conductance~\cite{Glazman88,Goldhaber98}.
Another remarkable effect is the Fano resonance physics intertwined
with Kondo physics~\cite{Hofstetter01,Bulka01}. Recent advances
in nanofabrication of Aharonov-Bohm (AB) devices made possible a realization
and a systematic study of such phenomena~\cite{ABexp}. 

For systems where the interaction is absent, or very weak, transport
properties may be determined using the Landauer-B\"{u}ttiker formalism~\cite{Landauer57}.
On the other hand, if the interaction is important in the system being
studied, the use of a more general approach is essential. An appropriate
formalism, expressing the conductance in terms of non-equilibrium
Green's functions, was developed by Meir and Wingreen~\cite{Meir92}.
The formalism can in principle be used to treat systems at a finite
temperature, finite source-drain voltage, and can even be extended
to describe time-dependent transport phenomena~\cite{Jauho94}. Another
approach to describing the transport in interacting systems, applicable
only to systems in the linear-response regime, is the Kubo formalism~\cite{Oguri}. 

Recently, a simple method for calculating the conductance through
a region with electron-electron interaction (e.g., a molecule, a quantum
dot, a quantum dot array or a similar 'artificial molecule' system,
...), connected to single-channel noninteracting leads, was presented
in Ref.~\onlinecite{Rejec03} (hereafter referred to as RR)~\cite{Favand98}.
The conductance is determined solely from the ground-state energy
of an auxiliary system, formed by connecting the ends of the leads
of the original system into a ring and threaded by magnetic flux.
The method is applicable to Fermi liquid systems at zero temperature
and in the linear response regime. The validity is additionally restricted
to the class of interacting systems obeying time reversal symmetry.
Several conductance formulae, becoming exact in the limit of a very
large ring, are derived, the most practical being the two-point formula,

\begin{equation}
G=G_{0}\sin^{2}\left(\frac{\pi}{2}\frac{E\left(\pi\right)-E\left(0\right)}{\Delta}\right),\label{eq:gsin}\end{equation}
where $E\left(0\right)$ and $E\left(\pi\right)$ are the ground-state
energies of the auxiliary interacting system with periodic and antiperiodic
boundary conditions, respectively. The average level spacing at the
Fermi energy $\Delta=[N\rho\left(\epsilon_{F}\right)]^{-1}$ is determined
by the density of states in an infinite lead $\rho\left(\epsilon\right)$
and the circumference of the ring $N$, and $G_{0}=\frac{2e^{2}}{h}$
is the conductance quantum. The corresponding ground-state wave function
can be chosen real, which simplifies the determination of the ground-state
energy by, for example, variational or density matrix renormalization
group approaches. 

\begin{figure}[t]
\begin{center}\includegraphics[%
  bb=210bp 0bp 610bp 652bp,
  clip,
  width=6cm,
  keepaspectratio]{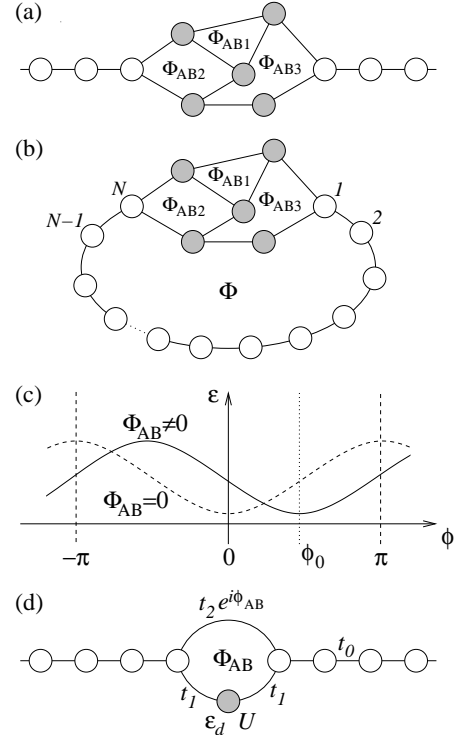}\end{center}

\caption{\label{cap:Fig1}(a) Interacting mesoscopic region (gray-shaded sites),
threaded by magnetic flux and coupled to noninteracting leads. (b)
Auxiliary ring system. (c) Behavior of energy levels as the flux threading
the ring is varied. (d) System from Ref.~\onlinecite{Hofstetter01}
with a quantum dot embedded in an AB ring.}
\end{figure}

In this paper we present a generalization of the conductance formula
Eq.~(\ref{eq:gsin}) to systems which explicitly exhibit time reversal
asymmetry, such as is e.g. an AB-type of system presented in Fig.~\ref{cap:Fig1}(a).
The formalism presented here is valid for general systems consisting
of interconnected sites with interaction, threaded with magnetic flux
and connected to noninteracting leads.

First we consider the noninteracting case and single-electron states.
The transmission amplitudes $t_{k}$ and $t_{k}^{\prime}$ of an electron
with a wave vector $k$ and an energy $\varepsilon_{k}$ describe
the transmission from the left to the right lead and from the right
to the left lead of the system presented in Fig.~\ref{cap:Fig1}(a),
respectively. They are related by an expression derived in RR,

\begin{equation}
t_{k}^{\prime}e^{i\phi}+t_{k}e^{-i\phi}=e^{ikN}+\frac{t_{k}}{t_{k}^{\prime\ast}}e^{-ikN},\label{eq:ttprime}\end{equation}
where $\Phi=\frac{\hbar}{e}\phi$ is the magnetic flux through the
auxiliary system shown in Fig.~\ref{cap:Fig1}(b). If there is no
AB flux threading the mesoscopic region, the time reversal symmetry
is restored, $t_{k}=t_{k}'$ and the energy is an even function of
$\phi$, as illustrated in Fig.~\ref{cap:Fig1}(c). In a general
case, the unitarity of the scattering matrix requires $\left|t_{k}\right|=\left|t_{k}^{\prime}\right|$
and the transmission amplitudes are related as $t_{k}=\tilde{t}_{k}e^{i\phi_{0k}}$
and $t_{k}'=\tilde{t}_{k}e^{-i\phi_{0k}}$. Expressing $\tilde{t}_{k}$
in terms of its modulus $\left|t_{k}\right|$ and a phase-shift $\varphi_{k}$,
$\tilde{t}_{k}=\left|t_{k}\right|e^{i\varphi_{k}}$, the eigenenergy
equation Eq.~(\ref{eq:ttprime}) reads\begin{equation}
\left|t_{k}\right|\cos(\phi-\phi_{0k})=\cos\left(kN-\varphi_{k}\right).\label{eigen}\end{equation}
Algebraic manipulation equivalent to that in RR gives an implicit
equation for $\left|t\left(\varepsilon_{k}\right)\right|$, exact
to the leading order in $\frac{1}{N}$, \begin{equation}
\frac{\partial\arccos\left(\mp\left|t\left(\varepsilon_{k}\right)\right|\cos\left[\phi-\phi_{0}\left(\varepsilon_{k}\right)\right]\right)}{\partial\cos\phi}=\pi N\rho(\varepsilon_{k})\frac{\partial\varepsilon_{k}}{\partial\cos\phi},\label{main}\end{equation}
where the sign $\mp$ depends on weather $k$ belongs to a decreasing
($+$) or to increasing ($-$) branch of the cosine function in Eq.~(\ref{eigen})
and $\rho(\varepsilon)=\bigl(\pi\sqrt{4t_{0}^{2}-\varepsilon^{2}}\bigr)^{-1}$
with $t_{0}$ being the nearest neighbor hopping in the leads. By
using the principle of mathematical induction, as in RR but generalized
to a finite $\phi_{0}$, the transmission probability at the Fermi
energy $\varepsilon_{F}$ is expressed as\begin{equation}
\frac{1}{\pi}\frac{\partial\arccos^{2}\left(\mp\left|t\left(\varepsilon_{F}\right)\right|\cos\left[\phi-\phi_{0}\left(\varepsilon_{F}\right)\right]\right)}{\partial\cos\phi}=\pi N\rho(\varepsilon_{F})\frac{\partial E}{\partial\cos\phi},\label{ground}\end{equation}
where $E$ is the ground-state energy of a system containing an even
number of electrons. Neglecting the variation of $t\left(\varepsilon_{F}\right)$
and $\rho\left(\varepsilon_{F}\right)$ with $\phi$ (the error made
is of the order of $\frac{1}{N}$), Eq.~(\ref{ground}) leads to
a universal form of the ground-state energy of the auxiliary system
$E(\phi)=\pi^{-2}\Delta\arccos^{2}\left(\mp\left|t\left(\varepsilon_{F}\right)\right|\cos\left[\phi-\phi_{0}\left(\varepsilon_{F}\right)\right]\right)+\mathrm{const.}$\cite{odd}
From this form, the transmission probability can be extracted, and
the conductance is given by\begin{equation}
G=G_{0}\sin^{2}\left(\frac{\pi}{2}\frac{E\left(\phi_{0}+\pi\right)-E\left(\phi_{0}\right)}{\Delta}\right),\label{eq:gsin2}\end{equation}
where $\phi_{0}\equiv\phi_{0}\left(\epsilon_{F}\right)$ is determined
by the position of the minimum (or maximum) in the energy vs. flux
$\phi$ curve, schematically shown in Fig.~\ref{cap:Fig1}(c). The
conductance can also be calculated from the more convenient four-point
formula~\cite{3point}:

\begin{alignat}{1}
G=G_{0}\biggl[ & \sin^{2}\left(\frac{\pi}{2}\frac{E\left(\pi\right)-E\left(0\right)}{\Delta}\right)+\nonumber \\
+ & \sin^{2}\left(\frac{\pi}{2}\frac{E\left(\pi/2\right)-E\left(-\pi/2\right)}{\Delta}\right)\biggr],\label{eq:gsin3}\end{alignat}
and $\phi_{0}$ can be determined from the expression\begin{equation}
\phi_{0}=-\arctan\frac{\sin\left(\frac{\pi}{2}\frac{E\left(\pi/2\right)-E\left(-\pi/2\right)}{\Delta}\right)}{\sin\left(\frac{\pi}{2}\frac{E\left(\pi\right)-E\left(0\right)}{\Delta}\right)}.\label{eq:phi0}\end{equation}

We now consider the interacting case. If no AB flux is present in
the mesoscopic region and the ground state of the system exhibits
Fermi liquid behavior, i.e., the perturbation theory in the interaction
strength is valid, the imaginary part of the self-energy due to the
interaction vanishes quadratically at the Fermi energy, \begin{equation}
\mathrm{Im}\Sigma_{ij}\left(\omega+i\delta\right)\propto\left(\omega-\varepsilon_{F}\right)^{2},\label{Luttinger}\end{equation}
and a quasiparticle Hamiltonian can be constructed for each value
of $\varepsilon_{F}$. As shown in RR the conductance calculated from
the quasiparticle Hamiltonian reproduces the conductance of the interacting
system, and is exactly given by Eq.~(\ref{eq:gsin}) {[}or Eq.~(\ref{eq:gsin3})
since $E\left(\pi/2\right)=E\left(-\pi/2\right)$ in this case{]}.
If the time reversal symmetry is broken due to AB flux, Eq.~(\ref{Luttinger})
is not valid and the proof has to be reconsidered. Repeating the steps
as presented in detail in RR, the proof is restored and basically
unchanged if the exact self energy obeys the relation\begin{equation}
\frac{1}{2i}\left[\Sigma_{ij}\left(\omega+i\delta\right)-\Sigma_{ij}\left(\omega-i\delta\right)\right]\propto\left(\omega-\varepsilon_{F}\right)^{2}.\label{Luttinger2}\end{equation}
It follows that also the linear response conductance of an interacting
AB system at zero-temperature is given by the four-point formula Eq.~(\ref{eq:gsin3}).
This condition is fulfilled if the system is a Fermi liquid. 

In order to demonstrate the practical value of the method, we quantitatively
analyze the conductance through an Aharonov-Bohm ring with a quantum
dot placed in one of the arms as presented in Fig.~\ref{cap:Fig1}(d)~\cite{Hofstetter01,Bulka01}.
The quantum dot is described as an Anderson impurity with level position
$\epsilon_{d}$ and a charging energy $U$, coupled to each of the
leads with a tunneling matrix element $t_{1}$. Electrons can also
be transferred from one lead to the other directly through the upper
arm of the AB ring. This process is described by a tunneling matrix
element $t_{2}$. The ring is threaded by an AB flux $\Phi_{AB}=\frac{\hbar}{e}\phi_{AB}$
in such a way that only the direct tunneling matrix element is affected,
i.e. $t_{2}\rightarrow t_{2}e^{i\phi_{AB}}$. The Fermi energy is
set at the middle of the band, thus $\Delta=\frac{2\pi t_{0}}{N}$.
We perform the finite-size analysis of the four-point formula Eq.~(\ref{eq:gsin3}),
changing the circumference of the ring $N$. In order to be able to
compare our results with those of the NRG method, we choose the same
values of parameters as in Ref.~\onlinecite{Hofstetter01}.

\begin{figure}[t]
\begin{center}\includegraphics[%
  clip,
  width=7cm,
  keepaspectratio]{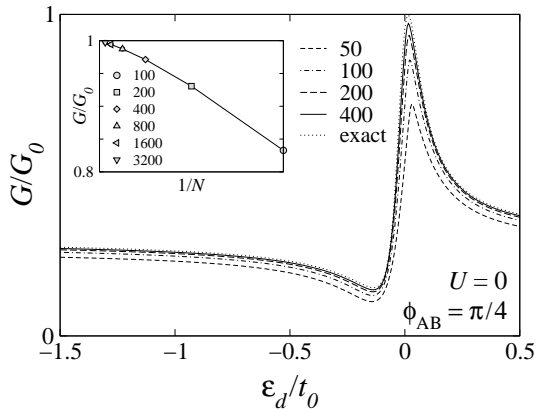}\end{center}

\caption{\label{cap:Fig2}The conductance of a noninteracting ($U=0$) AB
ring at $\phi_{AB}=\frac{\pi}{4}$. The inset shows convergence of
the conductance at the peak near $\varepsilon_{d}=0$. Parameters:
$t_{1}=0.177t_{0}$, $t_{2}=0.298t_{0}$.}
\end{figure}

In Fig.~\ref{cap:Fig2} the results for the conductance of a noninteracting
($U=0$) system with $\phi_{AB}=\frac{\pi}{4}$ are presented. Due
to a non-zero AB flux, the ground-state energy must be determined
at four $\phi$ points as required by Eq.~(\ref{eq:gsin3}). The
conductance exhibits a typical Fano resonance with a dip and a sharp
peak. Results calculated for various numbers of sites in the ring
$N$ are compared with the exact conductance curve. The inset shows
a convergence test of the method at the peak near $\epsilon_{d}=0$,
where due to a strong energy dependence of the transmission amplitude
the convergence is the most delicate.

In the interacting case, we determined the required ground-state energies
using an approach similar to the variational method of Gunnarson and
Sch\"{o}nhammer~\cite{Schonhammer75,Rejec03} for the Anderson model.
The variational basis set is generated from the ground state $\left|\tilde{0}\right\rangle $
of an auxiliary noninteracting Hamiltonian in which the energy level
of the dot $\varepsilon_{d}$ and the hopping matrix element $t_{1}$
between the dot and the leads are renormalized in such a way to minimize
the ground-state energy. Apart from the Hartree-Fock case where this
is the only variational wave function, we form two additional basis
sets. The first has three variational wave functions $P_{0}\left|\tilde{0}\right\rangle $,
$P_{1}\left|\tilde{0}\right\rangle $ and $P_{2}\left|\tilde{0}\right\rangle $,
where $P_{0}$, $P_{1}$ and $P_{2}$ are projectors onto unoccupied,
singly occupied and doubly occupied dot site. In the second basis
set we add four additional wave functions, $P_{0}VP_{1}\left|\tilde{0}\right\rangle $,
$P_{2}VP_{1}\left|\tilde{0}\right\rangle $, $P_{1}VP_{0}\left|\tilde{0}\right\rangle $
and $P_{1}VP_{2}\left|\tilde{0}\right\rangle $, where $V$ is the
operator describing hopping between the leads and the dot. In all
the three cases, the noninteracting limit is correctly reproduced.
Furthermore, except in the Hartree-Fock case, the method is also exact
in the limit where the dot is decoupled from the rest of the system.%
\begin{figure}[htbp]
\begin{center}\includegraphics[%
  clip,
  width=7cm,
  keepaspectratio]{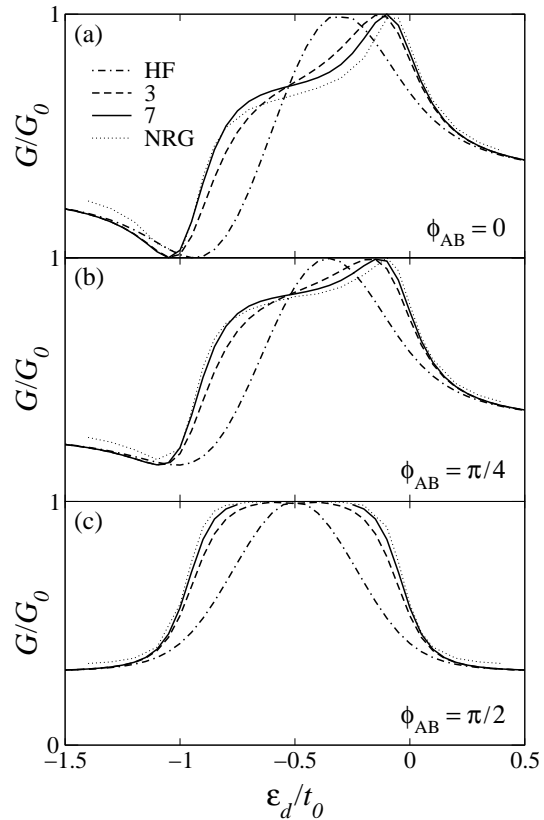}\end{center}

\caption{\label{cap:Fig3}Zero-temperature linear response conductance as
a function of level position $\varepsilon_{d}$ for various values
of the AB flux. Labels '3' and '7' correspond to variational wave
functions with three and seven terms, respectively. The dash-dotted
line is the result obtained using the HF approximation and the dotted
line is the NRG result from Ref.~\onlinecite{Hofstetter01}. Parameters:
$t_{1}=0.177t_{0}$ ($\Gamma=0.125t_{0}$), $t_{2}=0.298t_{0}$, $U=t_{0}=8\Gamma$.}
\end{figure}

An interacting AB ring with $U\neq0$ serves as a good nontrivial
test of the method. We choose a strong coupling regime as in Ref.~\onlinecite{Hofstetter01},
with $U=8\Gamma$, where $\Gamma=4\pi t{}_{1}^{2}\rho\left(\epsilon_{F}\right)$
is the line width of the dot level in the absence of the upper arm
of the AB ring. The results are presented in Fig.~\ref{cap:Fig3}
and compared with results obtained using the NRG method~\cite{Hofstetter01},
together with the corresponding Hartree-Fock (HF) curve. It should
be noted that with the parameter set chosen, the underlying physics
is in the strong correlation regime, where e.g. the width of the Kondo
peak is of the order $10^{-4}U$. The HF method therefore fails to
reproduce NRG results, while the variational results are very close
to the NRG curve. By extending the variational space from 3 to 7 terms,
the agreement with NRG improves even further and the small remaining
discrepancy would most probably be additionally reduced if an even
richer variational wave function is used. Interestingly, we find the
largest deviation in the 'empty orbital' regime, $\varepsilon_{d}\lesssim-U$.
The results of the variational and the HF methods here agree, both
deviating from the NRG result, while asymptotically all the curves
reach the correct limit $G=0.3G_{0}$~\cite{Hofstetter01}. We have
also checked the result in this regime using the second order perturbation
theory approach~\cite{Horvatic87} which gives a conductance in agreement
with variational and HF methods (not shown). 

\begin{figure}[t]
\begin{center}\includegraphics[%
  clip,
  width=7cm,
  keepaspectratio]{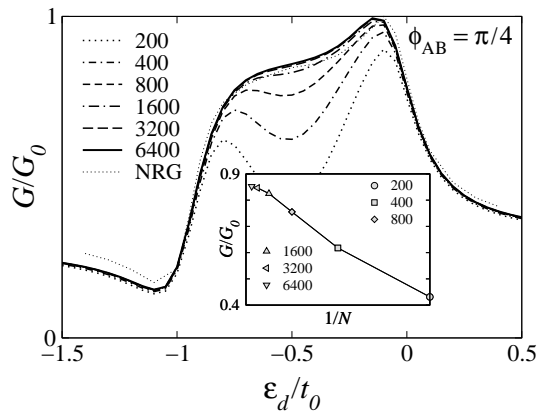}\end{center}

\caption{\label{cap:Fig4}The same as in Fig.~\ref{cap:Fig3}(b), but calculated
for different numbers of sites in the auxiliary ring $N$. The inset
shows convergence of the conductance at $\epsilon_{d}=-U/2$.}
\end{figure}

In Fig.~\ref{cap:Fig4} a convergence test of the method for the
result from Fig.~\ref{cap:Fig3}(b) is shown. The convergence with
$N$ is fast in the empty orbital regime and becomes progressively
slower as $\epsilon_{d}$ shifts toward the Kondo regime. The reason
can clearly be attributed to a very strong coupling regime resulting
in an extremely narrow Kondo peak. A very fine energy resolution is
therefore required to resolve the peak and to obtain a converged conductance
curve. 

\begin{figure}[b]
\begin{center}\includegraphics[%
  clip,
  width=7cm,
  keepaspectratio]{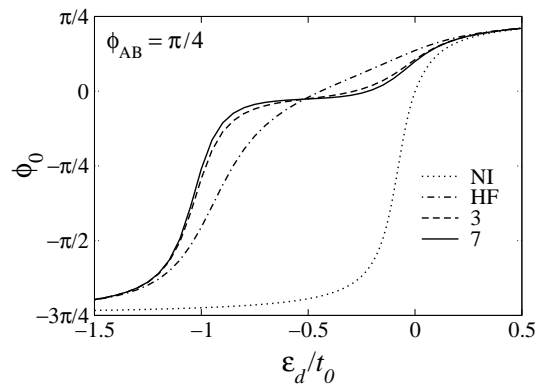}\end{center}

\caption{\label{cap:Fig5}Phase shift $\phi_{0}$ as a function of $\varepsilon_{d}/t_{0}$
for the noninteracting AB ring of Fig.~\ref{cap:Fig2} (NI) and for
various ground-state energy methods applied to the interacting AB
ring of Fig.~\ref{cap:Fig3}(b).}
\end{figure}

Broken time reversal symmetry in AB systems is signaled by $\phi_{0}\neq0$.
In Fig.~\ref{cap:Fig5} is shown the phase shift $\phi_{0}$ corresponding
to Figs.~\ref{cap:Fig2} and \ref{cap:Fig3}(b) as determined from
Eq.~(\ref{eq:phi0}). In contrast to the smooth noninteracting result,
the phase shift in the interacting case exhibits a well developed
plateau, corresponding to the Fano-suppressed Kondo plateau in the
conductance. The $\phi_{0}$ curve, which clearly cannot be correctly
reproduced in the HF approximation, can be explained as follows: In
the empty orbital regime, the current mainly flows through the upper
arm of the ring and therefore, electrons acquire an additional phase
shift $\phi_{0}\sim\phi_{AB}$ (note that $\phi_{0}$ and $\phi_{0}-\pi$
are physically equivalent). On the other hand, in the Kondo regime
almost all the current passes through the quantum dot and no additional
phase shift is present, $\phi_{0}\sim0$.

In summary, we have derived a formula for zero-temperature linear
response conductance of an interacting electron region coupled to
single-channel leads where the electron-electron interaction is absent.
The interacting part of the system can be a general Aharonov-Bohm
type of interferometer with broken time reversal symmetry {[}Fig.~\ref{cap:Fig1}(a){]}.
The conductance of such an \emph{open} system is exactly determined
from an auxiliary \emph{closed} system {[}Fig.~\ref{cap:Fig1}(b){]},
where the leads of the original system are connected to form a ring
of $N$ noninteracting sites threaded by an auxiliary magnetic flux.
Eq.~(\ref{eq:gsin3}), which follows from the universal form of the
ground-state energy of the auxiliary system for a large but finite
$N$, expresses the conductance in terms of the ground-state energy
evaluated at four different values of the auxiliary flux. The proof
of validity of the formula for interacting systems relies on the mapping
of the system onto an effective quasiparticle problem and is therefore
valid for systems exhibiting Fermi liquid properties. We have demonstrated
the usefulness of the formula by applying it to a prototype system
exhibiting Kondo-Fano behavior. Results based on the four-point formula
and variational ground-state energies confirm results of the numerical
renormalization group method.

The authors wish to acknowledge P. Prelov\v{s}ek and X. Zotos for helpful
discussions and W. Hofstetter for useful correspondence regarding
his results. We acknowledge J.~H. Jefferson for useful remarks and
the financial support of QinetiQ.


\begin{thebibliography}{10}
\bibitem{odd}For an odd number of electrons in a system, the universal form of
the ground-state energy is $E(\phi)=\pi^{-2}\Delta\arcsin^{2}\left(\left|t\left(\varepsilon_{F}\right)\right|\cos\left[\phi-\phi_{0}\left(\varepsilon_{F}\right)\right]\right)+\mathrm{const.}$
\bibitem{3point}It is also possible to calculate the conductance from the ground-state
energies at three distinct values of flux $\phi$ by solving Eq.~(\ref{ground})
numerically.
\end{thebibliography}
\end{document}